\documentclass[a4paper]{article}

\usepackage{INTERSPEECH2021}
\usepackage{multirow}
\usepackage{url}
\usepackage{arydshln}

\title{GIST-AiTeR Speaker Diarization System for VoxCeleb Speaker Recognition Challenge (VoxSRC) 2023}
\name{
    Dongkeon Park$^1$,
    Ji Won Kim$^1$, 
    Kang Ryeol Kim$^1$,
    Do Hyun Lee$^1$,
    and Hong Kook Kim$^{1,2}$
    }
\address{
    $^1$AI Graduate School,
    $^2$School of Electrical Engineering and Computer Science\\
    Gwangju Institute of Science and Technology, Gwangju 61005, Republic of Korea
}
\email{\{dongkeon@gm, jiwon.kim@gm., kr\_kim99@gm., zerolee12@gm., hongkook@\}gist.ac.kr}

\begin{document}

\maketitle

\begin{abstract} 
    This report describes the submission system by the GIST-AiTeR team for the VoxCeleb Speaker Recognition Challenge 2023 (VoxSRC-23) Track 4. Our submission system focuses on implementing diverse speaker diarization (SD) techniques, including ResNet293 and MFA-Conformer with different combinations of segment and hop length. Then, those models are combined into an ensemble model. The ResNet293 and MFA-Conformer models exhibited the diarization error rates (DERs) of 3.65\% and 3.83\% on VAL46, respectively. The submitted ensemble model provided a DER of 3.50\% on VAL46, and consequently, it achieved a DER of 4.88\% on the VoxSRC-23 test set. 
\end{abstract}


\section{Introduction} 
    Speaker diarization (SD) is a task of determining “who spoke when,” and it plays a vital role in many applications involving multi-speaker audio data.
Traditionally, SD is framed as a clustering problem where each segment of speech is assigned a particular speaker label.
A typical clustering-based SD system is modular in nature and comprises components for voice activity detection (VAD), speaker embedding extraction, and speaker clustering \cite{Shum2013}.
Once speaker labels are associated with each segment, segments sharing the same label are grouped together.
By improving speaker embedding, traditional SD systems have achieved remarkable accomplishments through substantial reduction of speaker ambiguity.
Recent efforts to refine this process have been concentrated on exploring methodologies such as overlapped speech detection (OSD) \cite{overlap4518619}, end-to-end neural SD (EEND) \cite{fujita2020end}, and target-speaker voice activity detection (TS-VAD) \cite{medennikov2020target} to mitigate missed speaker errors.

In this report, we propose a clustering-based SD system specifically tailored for the Diarization Task of the VoxCeleb Speaker Recognition Challenge 2023 (VoxSRC-23). The proposed approach integrates several processing stages such as VAD, speaker embedding extraction, clustering, and OSD.

The remainder of this report is organized as follows: Section 2 delves into the proposed system architecture, outlining four distinctive systems developed by different combinations of SD modules with various hyperparameter settings. Then, it also describes the construction of fusion systems to enhance diarization efficiency. Section 3 presents and discusses the experimental results. Finally, Section 4 concludes this report.
\label{sec:intro}

\section{GIST-AiTeR SD System}\label{sec:system}
    The following subsections will explain each SD module in detail. In accordance with \cite{wang2021dku}, the provided challenge dataset is divided into two distinct subsets: 
a development set (DEV402) and a validation set (VAL46) that are composed of the initial 402 recordings and the remaining 46 recordings, respectively.

%\begin{table}[ht]
%\begin{table}[htb]
\begin{table}[t]
  \caption{Comparison of Precision and F1-score of different OSD modules on VAL46}
  \label{tab:ovd}
  \vspace{-2mm}
  \centering
  \resizebox{\linewidth}{!}{
  \begin{tabular}[c]{lcc}
    \toprule
     \textbf{OSD Module} & \textbf{Precision} (\%)& \textbf{F1-score} (\%) \\
    \midrule
    GIST-AiTeR (VoxSRC-22) \cite{park2022gist}  & 88.81  & 52.45 \\
    ResNet+LSTM with Multitask detection      & \textbf{90.16}  & \textbf{53.81} \\
    %Fusion(1+2+3)       & \textbf{88.81} & 37.22 & 52.45 \\
    \bottomrule

    \vspace{-2mm}
  \end{tabular}
  }
\end{table}
    \subsection{VAD and OSD}\label{sec:vad}
        VAD identifies speech among audio segments. In this work, we utilized the identical approach as the one used in our VoxSRC-22 submission system \cite{park2022gist}. Specifically, we constructed an ensemble method with ResNet+LSTM and SincNet+LSTM modules, averaging their posterior predictions.

For OSD, we utilized the same ResNet+LSTM model structure and training dataset as those used in \cite{park2022gist}. In addition, multitask detection in \cite{2023MTL} was employed for activating speech, overlapping speech, and speaker changes. In particular, we used weighted binary cross entropy (BCE) loss to train for OSD owing to the data imbalance problem. This imbalance occurs because the number of overlapped speech frames is significantly smaller than that of single speech or silence frames. The weighted BCE loss was also used for speaker change detection, effectively addressing imbalance problem. To adjust the hyperparameters in our model, we utilized Pyannote’s hyperparameter optimization \cite{bredinpyannote}. In particular, we set the $\beta$ value to 0.1 in the $F_{\beta}$ score calculation, thereby increasing the importance of precision in the optimization process. This adjustment ensures accurate identification of overlap regions while prioritizing precision.

Table \ref{tab:ovd} compares the precision and F1-score of different OSD modules on VAL46.  As shown in the table, the ResNet+LSTM model with the multitask detection approach improved both precision and F1-score compared to our submission system from the previous year. Consequently, we selected this model as the OSD module for our proposed SD system.

\begin{table}[tp]
  \caption{Performance comparison of our different versions of SD systems}
  \vspace{-2mm}
  \label{tab:result}
\centering
{\small
\resizebox{\linewidth}{!}{
\begin{tabular}{lcccccc}
\toprule
\textbf{Method} & \textbf{Time-scale} & \multicolumn{2}{c}{\textbf{VAL46}} & \multicolumn{2}{c}{\textbf{VoxSRC-23}} \\
\cmidrule(lr){3-4} \cmidrule(lr){5-6}
 &  & \textbf{DER} & \textbf{JER} & \textbf{DER} & \textbf{JER} \\

\midrule
GIST-AiTeR \cite{park2022gist} & - & 3.56 & 27.63 & 5.67 & 33.51 \\
\midrule[0.01pt]
\multirow{3}{*}{MFA-Conformer} & 1.0 / 0.5 & 3.98 & 25.89 & - & - \\
 & 2.0 / 1.0 & 3.83 & 24.33 & - & - \\
 & 3.0 / 1.5 & 4.11 & 26.02 & - & - \\
\midrule[0.01pt]
ResNet293 & 1.5 / 0.25 & 3.65 & 24.97 & - & - \\
\midrule[0.01pt]
DOVER-Lap & & \textbf{3.50} & \textbf{25.23} & \textbf{4.88} & \textbf{29.69} \\
\bottomrule
\end{tabular}
}
}
\vspace{-2mm}
\end{table}

    \subsection{Speaker Embedding}\label{sec:xvec}
        In our speaker embedding process, we employed two models: MFA-Conformer, trained with a multiscale approach and consistent with the process described in the previous year’s implementation \cite{park2022gist}, and ResNet293 provided by Wespeaker \cite{wang2022wespeaker}. While MFA-Conformer was employed solely in the previous year, we introduced ResNet293 as an innovative component to complement MFA-Conformer.
The ensemble of the MFA-Conformer and ResNet293 forms the core of the proposed SD system, providing a robust solution to extract speaker embedding.

    \subsection{Clustering}\label{sec:clustering}
        The dimension of speaker embedding vectors obtained from both MFA-Conformer and ResNet293 was first reduced into 128 using the linear discriminant analysis (LDA). Subsequently, two probabilistic LDA (PLDA) models were formulated by utilizing the 128-dimensional embedding vectors, derived from both VoxCeleb1 and VoxCeleb2 and the VoxConverse DEV402 set. Then, these two PLDA models were interpolated into one with a factor of 0.9.

Next, agglomerative hierarchical clustering (AHC) was applied to a given utterance, where PLDA scores were employed to allocate a cluster index to individual frame. A threshold for AHC clustering was determined with a two-component Gaussian mixture model using a shared variance over all PLDA scores from the utterance.

After that, the variational Bayesian hidden Markov model (VB-HMM) \cite{landini2022bayesian} was applied to reassign the cluster index to each frame, considering time dependencies. This approach allows more freedom to find optimal results and converge to the appropriate number of speaker models.

    % \subsection{Overlapped Speech Detection}\label{sec:osd}
    %     \input{1_Index/2_5_OVD}

% \section{Dataset Description} \input{1_Index/3_Dataset}

\section{Experimental Results} 
    Table \ref{tab:result} compares the SD performance of different versions of our system, including our previous model from VoxSRC-22 \cite{park2022gist}, MFA-Conformer, ResNet293, and the final ensemble DOVER-Lap model. The performance was evaluated on both VAL46 and VoxSRC-23 test set. In this report, the performance of MFA-Conformer was evaluated across three different combinations of segment length and hop length, such as (1 s / 0.5 s), (2 s / 1 s), and (3 s / 1.5 s), with the hop length being half of the segment length. However, the ResNet293 embedding was evaluated using one pair of (1.5 s / 0.25 s). As shown in the table, different combinations of segment and hop lengths for MFA-Conformer led to varied performance results on VAL46, where a segment/hop length of (2.0 / 1.0) yielded the best diarization error rate (DER) of 3.83\%. On the other hand, the ResNet293 speaker embedding lowered DER by 0.18\% than MFA-Conformer, which was the best among all the single models evaluated in this work. 

Next, we ensembled various models according to different speaker embedding techniques alongside varying segment and hop lengths, through DOVER-Lap \cite{raj2021dover}, which culminated in a significantly reduced DER of 3.50\% on VAL46. This strategy of incorporating diverse modeling parameters played an important role in the success of the ensemble system, leading to its selection for the challenge. Consequently, the system obtained from this ensemble approach achieved a DER of 4.88\% on the challenge test set.

\section{Conclusion} 
    In this report, we explored various SD techniques and evaluated their performance on VAL46 and VoxSRC-23 test set. The ResNet293 and MFA-Conformer models exhibited varied performance depending on segment and hop lengths, resulting in the best DERs of 3.65\% and 3.83\%, respectively. By integrating all the individual models according to different combinations of segment and hop lengths through DOVER-Lap, we achieved a significantly reduced DER of 3.50\% on VAL46. Notably, this ensemble approach led to a DER of 4.88\% on the challenge test set, ranked fourth place in the VoxSRC-23.

\section{Acknowledgements} 
    % This research was supported by Culture, Sports, and Tourism R\&D Program through the Korea Creative Content Agency grant funded by the Ministry of Culture, Sports, and Tourism (R2022060001) in 2022, by Institute of Information \& Communications Technology Planning \& Evaluation (IITP) grant funded by the Korea government (MSIT) (No. 20220-00963), and by the Korean National Police Agency [Pol-Bot Development for Conversational Police Knowledge Services / PR09-01-000-20].

This work was supported in part by Institute of Information \& Communications
Technology Planning \& Evaluation funded by the Korea Government (MSIT) (2022-0-00963)
and by Culture, Sports, and Tourism R\&D Program through the Korea Creative Content Agency grant funded by the Ministry of Culture, Sports, and Tourism in 2023 (R2022060001, Contribution Rate: 50\%).

\bibliographystyle{IEEEtran}
\bibliography{main}

\end{document}